\documentclass[%
 ,twocolumn%
 ,secnumarabic%
,amssymb, amsmath,nobibnotes, prl, aps]{revtex4}
\usepackage{bm,epsf}%

\newcommand{\g}{{\gamma}}

\newcommand{\la}{{\lambda}}

\expandafter\ifx\csname package@font\endcsname\relax\else
 \expandafter\expandafter
 \expandafter\usepackage
 \expandafter\expandafter
 \expandafter{\csname package@font\endcsname}%
\fi

\begin{document} \draft

\title{Relativity Restored: Dirac Anisotropy in QED$_3$}
\author{O. Vafek$^1$, Z. Te\v{s}anovi\'c$^1$, and M. Franz$^2$}
\address{$^1$Department of Physics and Astronomy, Johns Hopkins
University, Baltimore, MD 21218, USA \\
$^2$Department of Physics and Astronomy, 
University of British Columbia, Vancouver, BC, Canada V6T 1Z
\\ {\rm(\today)}
}
\begin{abstract}
\medskip
We show that at long lengthscales and low energies and to leading order in 
$1/N$ expansion, the anisotropic QED
in 2+1 dimensions renormalizes to an isotropic limit. 
Consequently, the (Euclidean) relativistic invariance of the
theory is spontaneously restored at the isotropic critical point,
characterized by the anomalous dimension exponent of the Dirac fermion
propagator $\eta$. We find $\eta=16/3\pi^2 N$.
\end{abstract}

\maketitle

\narrowtext

Quantum electrodynamics in (2+1) dimensions (QED$_3$) has
recently emerged as a low-energy effective theory of a
number of condensed matter 
systems \cite{leewen,rantner,khvesh,ftqed,tvfqed,herbut,reenders}. 
Examples range from fluctuating d-wave superconductors in uderdoped cuprates 
\cite{ftqed,tvfqed,herbut}
to pyrolitic graphite \cite{khvesh} to
Heisenberg antiferromagnets and spin liquids \cite{leewen,rantner,reenders}.
While these multiple reincarnations
of QED$_3$ differ mightily in their physical content, they
all share certain important formal
similarities. The low energy behavior is controlled by an infra-red 
fixed point where the gauge field acquires a universal dimensionless
coupling constant $g\propto 1/N$, $N$ being the
number of Dirac fermion flavors \cite{appelquist,gusynin}. 
At values of $g$ larger than some critical value $g_c$ ($N<N_c$)
it is believed that the theory has an 
instability into a state with broken 
chiral symmetry, with fermions spontaneously acquiring 
a finite dynamical mass \cite{appelquist}.
Among the formal aspects shared by the above theories surely one of
the most ubiquitous is the spacetime anisotropy -- such
low-energy effective theories are obviously only pretending
to be ``relativistic''. They hail from non-relativistic quantum
Hamiltonians and are not obliged to be invariant under 
Lorentz transformations. Consequently, they
often contain more than one ``speed of light'' resulting in
the above anisotropy. An important question is to what
extent are the properties of these effective theories 
similar to the genuine, isotropic QED$_3$ and, in particular,
what is the nature of the critical behavior and chiral
symmetry breaking when such anisotropy is present.

In this Letter we address the problem of Dirac anisotropy
in QED$_3$. Our point of departure  is the assumption that the
{\em symmetric} (massless or critical) phase of 
{\em isotropic} QED$_3$, obtained
when the number of fermion flavors is larger than a critical
value, $N>N_c$, is controlled by a stable non-trivial infrared
critical point \cite{appelquist,gusynin} 
characterized by the anomalous dimension 
exponent $\eta >0$ of the {\em gauge-invariant} Dirac 
fermion propagator. 
Based on this assumption, which is almost certainly correct
for $N\gg N_c$, we derive the following results within a $1/N$ 
expansion: {\em i)}~When
{\em anisotropy} is turned on at this interacting critical
point we find it to be {\em marginally irrelevant} in a 
perturbative sense.
This implies that the symmetric, critical phase of QED$_3$ remains
unaffected by small Dirac anisotropy. In particular,
the value of $\eta$ remains unchanged. {\em ii)}~Going
beyond the perturbative regime, and by exploring the
structure of renormalization group (RG) flows,
we argue that any {\em finite}
anisotropy is also irrelevant. Finally, {\em iii)}~we
compute the explicit value of $\eta$ and find $\eta = 16/3\pi^2 N$.  
Our results imply that the relativistic 
invariance of a QED$_3$-like effective
theory is itself an {\em emergent} property: it is spontaneously 
dynamically restored at the critical point.  

The anisotropic QED$_3$ can be defined as follows:
\begin{equation}\label{lagrangian}
{\cal L}= 
\bar{\psi}^{(n)}[ \gamma_{\nu} \sqrt{g^{(n)}}_{\mu\nu}
(\partial_\nu+i a_{\nu})] \psi^{(n)}+\frac{1}{2e^2}(\partial\times a)^2
\end{equation}
where $\psi^{(n)}$ is a Fermi field associated with a node $n$, 
$\g_{\mu}$ is a Dirac matrix, and $a_{\mu}$ is 
a massless $U(1)$ gauge field related to fluctuations
of unbound $2+1$ vortex loops\cite{ftqed,tvfqed}.
We also introduced the diagonal ``nodal'' metric $g^{(n)}_{\mu\nu}$:
$g^{(1)}_{00}\!=\!g^{(2)}_{00}\!=\!1$,
$g^{(1)}_{11}\!=\!g^{(2)}_{22}\!=\!v^2_F$,
$g^{(1)}_{22}\!=\!g^{(2)}_{11}\!=\!v^2_{\Delta}$.
Other forms can be reduced to this one by suitable rescalings of
spacetime coordinates and fermion and gauge fields. 

The Dirac anisotropy of ${\cal L}$ (\ref{lagrangian}) is more sinister that
its bosonic kin \cite{herbutzbt}.
In the Higgs-Abelian gauge theory the anisotropy can be
fully rescaled out of the matter part of the action
leading to the new effective action with the anisotropy
stored only in the gauge field Maxwellian action.
Since matter cannot generate any anisotropic contribution
to the gauge field and by the virtue of the isotropic
charge being a relevant operator, it is easy to show that
anisotropy of the bosonic theory is {\em maginally irrelevant}.

In the fermionic QED$_3$ (\ref{lagrangian}) the above 
simple procedure does not work because one cannot simultaneously 
rescale the kinetic
energy for all fermion species. We therefore keep the anisotropy
confined to the matter part of (\ref{lagrangian}) and proceed
from there.
The two-point vertex function of the non-interacting theory for, say,
$(1,\bar{1})$ Dirac fermions is
\begin{equation}\label{2ptfreevertex}
\Gamma^{(2)free}_{1\bar{1}}= \gamma_0 k_0 +
v_F \gamma_1 k_1 + v_{\Delta} \gamma_2 k_2
\end{equation}
and the corresponding Green function equals
\begin{equation}
G^n_0(k)=\frac{\sqrt{g^n}_{\mu\nu}\gamma_{\mu}k_{\nu}}{k_{\mu}g_{\mu\nu}k_{\nu}} \equiv
\frac{\gamma^n_{\mu}k_{\mu}}{k_{\mu}g_{\mu\nu}k_{\nu}}~~.
\end{equation}

In what follows 
we assume that both $v_F$ and $v_{\Delta}$ are dimensionless and that
eventually one of them can be chosen to be unity by the appropriate
choice for the "speed of light". The anisotropy parameter
$\alpha_D = v_F/v_\Delta \neq 1$ breaks the Lorentz invariance of the theory 
(\ref{lagrangian}). However, the theory still respects 
time-reversal and parity and 
for $N>N_c$ the system is in its 
chirally symmetric phase \cite{leeherbut}. 
These symmetries force the fermion self-energy of the {\em interacting}
theory to assume the following form:
\begin{equation}
\Sigma_{1\bar{1}} =A(k) \left(\gamma_0 k_0 + v_F\zeta_1 \gamma_1 k_1 +
v_{\Delta} \zeta_2 \gamma_2 k_2 \right)~~,
\end{equation}
The coefficients $\zeta_i$ are in general different from unity.
Furthermore, there is a discrete spatial symmetry 
which relates flavors $(1,\bar{1})$ and $(2,\bar{2})$
to the $x$ and $y$ directions in such a way that 
\begin{equation}
\Sigma_{2\bar{2}} =A'(k) \left(\gamma_0 k_0 + v_{\Delta} \zeta_2 \gamma_1 k_1 
+ v_F \zeta_1 \gamma_2 k_2 \right)~~.
\end{equation}
In the computation of the fermion self-energy, 
this discrete symmetry allows us to 
concentrate on a particular pair of nodes
without any loss of generality.  

Next, we turn to the gauge field propagator. We first work in
the Lorentz gauge ($k_{\mu}a_{\mu}=0$) and then extend 
our results to a general covariant gauge.
To one-loop order the fermionic ``screening'' of the gauge
field is given by the polarization function 
\begin{equation}
\Pi_{\mu\nu}(k)=\frac{N}{2}\sum_{n=1,2} \int \frac{d^3q}{(2\pi)^3}Tr[G_0^n(q)\gamma_{\mu}
^n G_0^n(q+k)\gamma_{\nu}^n].
\end{equation}
The above expression can be evaluated
by observing that it reduces to the isotropic $\Pi_{\mu\nu}(k)$ 
once the integrals are properly rescaled \cite{ftqed}.
The result is:
\begin{equation}\label{polarization}
\Pi_{\mu\nu}(k)=\sum_n
\frac{N}{16v_Fv_{\Delta}}\sqrt{k_{\alpha}g^n_{\alpha \beta}k_{\beta}}
\left(
g^n_{\mu\nu}-\frac{g^n_{\mu\rho}k_{\rho}g^n_{\nu\lambda}k_{\lambda}}
{k_{\alpha}g^n_{\alpha \beta}k_{\beta}}
\right),
\end{equation}
where we have taken the advantage of the ``nodal'' metric $g^n_{\mu\nu}$
(\ref{lagrangian}).
This expression is explicitly transverse, i.e. 
$k_{\mu}\Pi_{\mu\nu}(k)=\Pi_{\mu\nu}(k)k_{\nu}=0$ and symmetric in its
spacetime indices. It also properly reduces to the isotropic expression
when $v_F=v_{\Delta}=1$. 

As opposed to the isotropic case, it is not quite as straightforward
to determine the gauge field propagator $D_{\mu\nu}$. 
To proceed we first integrate
out the fermions and expand the effective action to 
one-loop order
\begin{equation}\label{la}
{\cal L}_{\rm eff}[a_{\mu}]=(\Pi^{(0)}_{\mu\nu}+\Pi_{\mu\nu})
a_{\mu}(k)a_{\nu}(-k)~~,
\end{equation}
where the bare gauge field stiffness is
\begin{equation}
\Pi^{(0)}_{\mu\nu}=\frac{1}{2e^2} k^2 \left(\delta_{\mu\nu}-
\frac{k_{\mu}k_{\nu}}{k^2} \right)~~.
\end{equation}
Now we introduce the dual field 
$b_{\mu}=\epsilon_{\mu\nu\lambda}q_{\nu}a_{\lambda}$, which is
related to the physical fluctuating vorticity 
in the theory of Ref. \cite{ftqed}. 
We are free to integrate
over $b_\mu$ with the restriction that it
is transverse ($k_\mu b_\mu =0$). 
Note that 
\begin{equation}\label{lb}
{\cal L}_{\rm eff}[b_{\mu}]=\chi_0 b_0^2 +\chi_1 b_1^2 +\chi_2 b_2^2~~,
\end{equation} 
where $\{ \chi_{\mu}\}$ are functions of $k_{\mu}$:
\begin{equation}
\chi_{\mu}=\frac{1}{2e^2}+
\frac{N}{16v_Fv_{\Delta}}\sum_{n=1,2}
\frac{g^n_{\nu\nu}g^n_{\la\la}}{\sqrt{k_{\alpha}g^n_{\alpha\beta}k_{\beta}}}; \;\;
\mu\neq \nu\neq \lambda\in \{0,1,2\}.
\end{equation}
At low energies we can neglect the non-divergent bare stiffness and thus 
we set $1/e^2 =0$ in the above expression.

The expression (\ref{lb}) is manifestly gauge invariant and
has the merit of not only being 
quadratic but also diagonal in the individual components of $b_{\mu}$ which
greatly simplifies the computation of the  
$b_{\mu}$ correlation function: 
\begin{equation}
\langle b_{\mu} b_{\nu} \rangle=\frac{\delta_{\mu\nu}}{\chi_{\mu}}-
\frac{k_{\mu}k_{\nu}}{\chi_{\mu}\chi_{\nu}}\left( \sum_i \frac{k_i^2}{\chi_i}
\right)^{-1}.
\end{equation} 
The repeated indices are not summed over in the above expression. 

After this little trick with the integration over $b_\mu$,
it is now quite simple to compute the propagator for the 
original gauge field $a_{\mu}$ and in the Lorentz gauge we obtain:
\begin{equation}
D_{\mu\nu}(q)=\langle a_{\mu} a_{\nu} \rangle=
\epsilon_{\mu\alpha\beta }\epsilon_{\nu\lambda\rho} 
\frac{q_\alpha q_\lambda}{q^4}\langle b_\beta b_\rho \rangle~~.
\end{equation}
By employing the transverse character of $\langle b_{\mu}b_{\nu} \rangle$ 
(which is independent of the gauge)
the above expression can be further reduced to
\begin{equation}\label{bp}
D_{\mu\nu}(q)=
\frac{1}{q^2}\left(\left(\delta_{\mu\nu}-\frac{q_{\mu}q_{\nu}}{q^2}\right)
\langle b^2 \rangle - \langle b_{\mu} b_{\nu} \rangle \right)~~.
\end{equation} 

We use the above result to define a general ``covariant'' gauge 
for the anisotropic theory as
\begin{equation}\label{bpgi}
D_{\mu\nu}(q)=
\frac{1}{q^2}\left(\left(\delta_{\mu\nu}-(1-\frac{\xi}{2})\frac{q_{\mu}q_{\nu}}{q^2}\right)
\langle b^2 \rangle - \langle b_{\mu} b_{\nu} \rangle \right).
\end{equation}
where $\xi$ is a continuous gauge fixing parameter.
This expression is justified by the Fadeev-Popov procedure 
applied to the Lagrangian 
\begin{equation}\label{lxi}
{\cal L}_{\rm eff}[a_{\mu}]=\left(\Pi_{\mu \nu}+\frac{1}{\xi}\frac{2k^2}{\langle b^2 \rangle}
\frac{k_{\mu}k_{\nu}}{k^2}\right)a_{\mu}(k) a_{\nu}(-k)~~.
\end{equation} 
Note that $\langle b^2 \rangle$ can be 
determined without ever considering
the gauge fixing terms. The expression 
(\ref{bpgi}) is our final result for the gauge field propagator in an
anisotropic ``covariant'' gauge.

Having determined the free fermion and screened gauge field
propagators of the anisotropic theory, we can now compute
the Dirac fermion self-energy generated by the photon
exchange to leading order in $1/N$:
\begin{equation}
\Sigma_{n}(q)=\int \frac{d^3k}{(2\pi)^3}\gamma^{n}_{\mu}
G_0^{n}(q-k) \gamma^{n}_{\nu}D_{\mu\nu}(k)~~,
\end{equation}
where $n$ is the node index. After some tedious algebra this
can be manipulated into:
\begin{equation}\label{selfenergy}
\Sigma_{n}(q)=-\sum_{\mu} \eta^n_{\mu}(\g^n_{\mu}q_{\mu})
\ln{\left(\frac{\Lambda}{\sqrt{q_{\alpha}g^n_{\alpha\beta}q_{\beta}}}\right)}~~.
\end{equation}
Here $\Lambda$ is the ultraviolet cutoff and 
the coefficients $\eta_\mu$ are functions of the bare anisotropy which can 
be reduced to quadratures. In case of weak 
anisotropy ($v_F=1+\delta ,\;v_{\Delta}=1$) to second order in $\delta$:
\begin{equation}\label{eta0}
\eta_0^{1\bar1}=-\frac{8}{3\pi^2N}\left(1-\frac{3}{2}\xi-\frac{1}{35}\left(40-7\xi\right)\delta^2\right)
\end{equation}
\begin{equation}\label{eta1}
\eta_1^{1\bar1}=-\frac{8}{3\pi^2N}\left(1-\frac{3}{2}\xi+\frac{6}{5}\delta -\frac{1}{35}\left(43-7\xi\right)\delta^2\right)
\end{equation}
\begin{equation}\label{eta2}
\eta_2^{1\bar1}=-\frac{8}{3\pi^2N}\left(1-\frac{3}{2}\xi-\frac{6}{5}\delta-\frac{1}{35}\left(1-7\xi\right)\delta^2\right).
\end{equation}
In the isotropic limit ($v_F=v_{\Delta}=1$) 
we regain $\eta_{\mu}^n=-8(1-\frac{3}{2}\xi)/3\pi^2 N$ 
as previously found by others \cite{appelquist,gusynin}.

We are now in position to turn to our main concern: the 
effect of anisotropy at the above critical point of
isotropic QED$_3$.
Before plunging into formal analysis, we first make some general physical
observations regarding the RG flow of the anisotropy. 
First, by examining the Eq. (\ref{selfenergy}) it is clear
that if $\eta^n_{1}=\eta^n_{2}$
then the anisotropy does not flow and 
remains equal to its bare value. 
This would imply that anisotropy is marginal and the theory
is described by some anisotropic fixed point. 
For this to happen, however, there
would have to be a symmetry that preserves the equality
$\eta^n_1=\eta^n_{2}$. In the isotropic QED$_3$ 
the symmetry that protects the equality of $\eta_\mu$'s 
is the (Euclidean) Lorentz invariance. In our anisotropic
case this symmetry is broken and thus we generically have 
$\eta^n_1 \not= \eta^n_2$. Therefore, the anisotropy {\em runs}
in the RG sense and {\em flows away} from its bare value.
If we start with $\alpha_D>1$ and find that $\eta_2^{1\bar1}>\eta_1^{1\bar1}$
at some scale $p<\Lambda$, the anisotropy 
is {\em marginally irrelevant} and decreases towards unity as
we move toward infrared. 
On the other hand, if $\eta_2^{1\bar1} < \eta_1^{1\bar1}$, 
then the anisotropy 
becomes {\em marginally relevant},
continues increasing beyond its bare value and the theory ultimately
flows into some {\em new}, {\em anisotropic} critical point.

The above arguments concerning  $\eta_2^n-\eta_1^n$ and RG flows
are on solid ground physically only if they can be
made in a gauge-independent way. This condition appears compromised
by the fact that $\eta^n_{\mu}$'s are
gauge dependent quantities and explicitly include
the gauge fixing parameter $\xi$. However, the difference $\eta_2^n-\eta_1^n$
{\em is} itself {\em gauge-invariant}.
This is seen directly from
Eqs. (\ref{eta0}-\ref{eta2}) where the $\xi$ dependence of all $\eta$'s 
is exactly the same. This cancellation of $\xi$ dependent
terms in  $\eta_2^n-\eta_1^n$ occurs not only for $\delta\ll 1$ but is
the general feature of $\eta^n_{\mu}$'s to all orders
in anisotropy and for any choice of ``covariant'' gauge fixing. 
Therefore, the RG analysis that follows is fully gauge
invariant as it should be.

The renormalized two-point vertex function is related to the 
``bare'' vertex via a fermion field rescaling factor
$Z_{\psi}$ as $\Gamma^{(2)}_R=Z_{\psi}\Gamma^{(2)}.$
It is natural to demand that 
at some renormalization scale $p$,
$\Gamma^{(2)}_R(p)$ have the form
(at nodes $1$ and $\bar1$):
$\Gamma^{(2)}_R(p)=\g_0p_0+v_F^R\g_1p_1 + v_{\Delta}^R\g_2p_2$,
where $v_F^R$ and $v_\Delta^R$ are the renormalized velocities.
The above equation corresponds to our renormalization condition through 
which we can eliminate the cutoff dependence and compute the RG flows.

\begin{figure}
\epsfxsize=8.0cm
\hfil\epsfbox{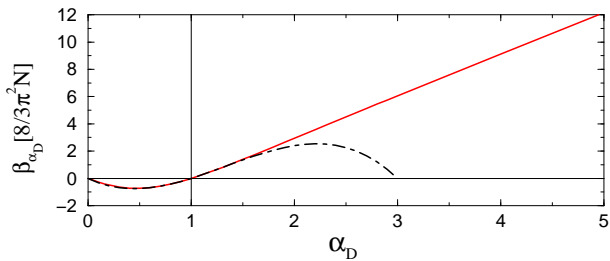}\hfill
\vspace{-0.25cm}
\caption{The RG $\beta$-function for the Dirac anisotropy in units of 
$8/3\pi^2N$. The solid line is the numerical integration
while the dash-dotted 
line is the analytical expansion around the small anisotropy 
(see Eq. (\ref{eta0}-\ref{eta2})). 
At $\alpha_D=1$, $\beta_{\alpha_D}$ crosses zero with positive slope, 
and therefore at large 
lengthscales the anisotropic QED$_3$ scales to an isotropic theory.}
\vspace{-.2cm}
\label{beta}
\end{figure}

To order $1/N$ we can write
\begin{equation}
\Gamma^{(2)}_R(p)=Z_{\psi}
\g^n_{\mu}p_{\mu}\left(1+\eta^n_{\mu}\ln{\frac{\Lambda}{p}}\right)~~,
\end{equation}
where we have used the self-energy (\ref{selfenergy}).
Multiplying both sides by $\g_0$ and taking the trace determines
the field strength renormalization:
\begin{equation}
Z_{\psi}=\frac{1}{1+\eta^n_0\ln{\frac{\Lambda}{p}}}\approx 
1-\eta^n_0\ln{\frac{\Lambda}{p}}~~.
\end{equation} 
We can now determine the renormalized Fermi and gap velocities:
\begin{equation}
\frac{v_F^R}{v_F}\approx
(1-\eta^{1\bar{1}}_0\ln{\frac{\Lambda}{p}})(1+\eta^{1\bar{1}}_1
\ln{\frac{\Lambda}{p}})\approx 
1-(\eta^{1\bar{1}}_0-\eta^{1\bar{1}}_1)\ln{\frac{\Lambda}{p}}
\end{equation}
and 
\begin{equation}
\frac{v_{\Delta}^R}{v_{\Delta}}\approx
(1-\eta^{1\bar{1}}_0\ln{\frac{\Lambda}{p}})(1+\eta^{1\bar{1}}_2
\ln{\frac{\Lambda}{p}})\approx
1-(\eta^{1\bar{1}}_0-\eta^{1\bar{1}}_2)\ln{\frac{\Lambda}{p}}.
\end{equation}
The corresponding renormalized Dirac anisotropy is therefore
\begin{equation}
\alpha_D^R\equiv\frac{v_F^R}{v_{\Delta}^R}\approx
\alpha_D\left[1-(\eta^{1\bar{1}}_2-\eta^{1\bar{1}}_1)
\ln{\frac{\Lambda}{p}}\right]~~.
\end{equation}
The RG beta function for the anisotropy is given by:
\begin{equation}
\beta_{\alpha_D}=\frac{d \alpha^R_D}{d\ln{p}}=
\alpha_D(\eta^{1\bar{1}}_2-\eta^{1\bar{1}}_1)~~.
\end{equation}
In the case of weak anisotropy ($v_F=1+\delta,\;v_{\Delta}=1$) the above 
expression can be determined analytically as an expansion in $\delta.$ 
Using  Eqs.(\ref{eta1}-\ref{eta2}) we obtain
\begin{equation}
\beta_{\alpha_D}=\frac{8}{3\pi^2N}\left(\frac{6}{5}\delta (1
+\delta)(2 -\delta)+{\cal O}(\delta^3)
\right).
\end{equation}
Note that this expression is independent of the gauge parameter $\xi$.
For $0<\delta\ll 1$ the $\beta$ function is positive 
which means that anisotropy 
decreases in the IR and thus
the anisotropic QED$_3$ scales to an isotropic QED$_3$.
For $-1\ll \delta <0$ the $\beta$ function is negative and in this case
$\alpha_D$ increases towards the fixed point $\alpha_D=1$, i.e. again
towards the isotropic QED$_3$. Note that for $\delta>2$, $\beta<0$  which may
naively indicate that there is a fixed point at $\delta=2$; this however 
cannot be trusted as it is outside 
of the range of validity of the power expansion of $\eta_{\mu}$. 
The numerical 
evaluation of the quadrature for $\eta_{\mu}$ shows that, apart from the 
isotropic fixed point and the unstable fixed point at $\alpha_D=0$,  
$\beta_{\alpha_D}$ does not vanish 
(see Fig. \ref{beta}). 
This indicates that to the leading order in $1/N$ expansion, 
the theory flows into the isotropic fixed point and
relativistic invariance is dynamically restored in the IR limit.

Although $\Sigma_n (q)$ is a gauge dependent quantity we 
now show that it can still be put to good use in helping us extract
the physical, gauge-invariant information about fermion propagation.
To this end, we employ an important
result derived by Brown \cite{brown} in the context of QED$_4$ which relates 
$G(x-x')=\langle \psi (x)\bar\psi (x')\rangle$ 
to the {\em gauge-invariant}
propagator $\tilde G(x-x')$:
\begin{equation}\label{giprop}
G(x-x')=e^{-F(x-x')} \tilde{G}(x-x')
\end{equation}
where 
\begin{equation}\label{yennie}
F=\frac{1}{2}\int dz dz' J_{\mu}(z)D_{\mu \nu}(z-z')J_{\nu}(z')
\end{equation}
and 
\begin{equation}\label{source}
J_{\mu}(z)=(x-x')_{\mu}\int_0^1 d\alpha \; \delta^d(z-x'-\alpha(x-x'))~.
\end{equation}
The exponential factor in (\ref{giprop}) is just
the expectation value of the straight line integral of the gauge field
$\langle\exp(i\int^{x'}_x ds_\mu a_\mu)\rangle$ averaged over (\ref{lxi}).

The expressions (\ref{giprop}-\ref{source}) 
hold in an arbitrary covariant
gauge and reflect the fact that
the part of ${\cal L}_{\rm eff}[a_\mu]$  
obtained after integrating out the fermions
must be purely transverse and thus 
the longitudinal, gauge dependent part (\ref{lxi}),  
enters only at the quadratic order. 
The leading long wavelength dependence is then given by 
Eqs. (\ref{giprop}-\ref{source}).
Using the long distance scaling of the gauge field propagator
$D_{\mu \nu}(s(x-x'))=s^{1-d} D_{\mu \nu}(x-x')$
and dimensional regularization \cite{brown} 
it is straightforward to show that
$F(r) = \frac{1}{(d-2)(d-3)}r_{\mu}D_{\mu \nu}(r)r_{\nu}~,$
where $D_{\mu \nu}(r)$ is the Fourier transform
of (\ref{bpgi}) to the real spacetime. 
Computing this expression at the isotropic fixed point
in the limit $d \rightarrow 3$ we find
\begin{equation}
F(r)=\frac{4(\xi-2)}{N\pi^2}\left( \frac{r^{3-d}}{3-d}\right)
\rightarrow \frac{4(\xi-2)}{N\pi^2}\left(\log(\Lambda r)+ \frac{1}{3-d}\right)
\end{equation}

If $\xi=2$ (Yennie gauge) $F$ {\em vanishes}
and the physical gauge-invariant
propagator $\tilde G(x-x')$ {\em coincides} with the gauge-variant 
one $G(x-x')$.
An important observation follows from the above
results: by inserting $\xi =2$ into our expression for the
gauge-variant self-energy (\ref{selfenergy}-\ref{eta2})
we obtain the anomalous dimension
exponent of the physical gauge-invariant Dirac fermion
propagator in isotropic QED$_3$, $\eta=16/3\pi^2N$ 
($\eta \sim 0.27$ for $N=2$) \cite{ftqed,gusynin2}. 
This exponent serves as the signature of
the interacting infrared critical point
which regulates the low-energy physics in the chiral
symmetric phase (AFL of Ref. \cite{ftqed}) of both
isotropic {\em and} anisotropic QED$_3$.

We thank V. Gusynin for sharing his valuable insights on QED$_3$ and 
Profs. I. F. Herbut and  D. V. Khveshchenko for useful comments.
This work was supported in part by 
NSF grant DMR00-94981 (OV and ZT) and
NSERC (MF).

\vspace{-0.5cm}


\begin{references}
\bibitem{leewen} D. H. Kim, P. A. Lee and X.-G. Wen, \prl {\bf 79}, 
2109 (1997).  
\bibitem{rantner} W. Rantner and X.-G. Wen, \prl {\bf 86}, 3871 (2001); 
W. Rantner and X.-G. Wen, cond-mat/0010378.
\bibitem{khvesh} D. V. Khveshchenko, \prl 87, 206401 (2001); see also 
cond-mat/0112202.
\bibitem{ftqed} M. Franz and Z. Te\v sanovi\' c, \prl {\bf 87}, 
257003 (2001).
\bibitem{tvfqed} Z. Te\v sanovi\' c, O. Vafek and M. Franz, cond-mat/0110253.
\bibitem{herbut} I. F. Herbut, cond-mat/0110188.
\bibitem{reenders} M. Reenders, cond-mat/0110168.
\bibitem{appelquist}R. D. Pisarski, \prd {\bf 29}, 2423 (1984);
T. W. Appelquist {\em et al.}, \prd {\bf 33},
3704 (1986); M. R. Pennington and D. Walsh, Phys. Lett. B {\bf 253}, 246
(1991); I. J. R. Aitchison and N. E. Mavromatos, \prb {\bf 53},
9321 (1996); P. Maris, \prd {\bf 54}, 4049 (1996). 
\bibitem{gusynin} V. Gusynin, A. Hams and M. Reenders, 
\prd {\bf 63}, 045025 (2001).
\bibitem{herbutzbt} I. F. Herbut and Z. Te\v sanovi\' c, \prl {\bf 51}, 
16204 (1995); Z. Te\v sanovi\' c, \prb {\bf 59}, 6449 (1999).  
\bibitem{leeherbut} Note that $N_c$ itself could depend on anisotropy;
see D. Lee and I. F. Herbut, cond-mat/0201088.
\bibitem{ftvqed} M. Franz, Z. Te\v sanovi\' c and O. Vafek, in preparation.
\bibitem{brown} L. S. Brown, {\em Quantum Field Theory}, 
(Cambridge University Press, 1992).
\bibitem{gusynin2} This exponent has also been computed by V. Gusynin,
private communication.

\end{references}
\end{document}